# Self-Organized Criticality in Asynchronously Tuned Elementary Cellular Automata

Yukio-Pegio Gunji[1, 2]

[1]*Department of Earth & Planetary Science, Faculty of Science, Kobe University, Nada Kobe 657-8501, Japan*
[2]*The Unconventional Computing Centre, University of the West England, Bristol, BS16 1QY, UK*

Self-organized criticality (SOC) reveals a mechanism by which a system is autonomously evolved to be in a critical state without needing parameter tuning. Whereas various biological systems are found to be in critical states and the significance of SOC is being re-estimated, a simple model in a general platform has not been established. Here, we present SOC in asynchronously tuned elementary cellular automata (ECA), which was based on asynchronously updating and tuning the consistency between local dual modes of transitions. This duality was defined by adjunction, which can be ignored during synchronous updates. Duality coupled with asynchronous updating can demonstrate that SOC coincides with the criticality in a phase transition of asynchronous ECA with respect to density decay.

Since Bak and his colleagues proposed the idea of self-organized criticality (SOC) [1-3], the importance of SOC has increased, particularly in biological systems [4-6]. It has been found that various biological systems might be at the edge of chaos [7-10], which can be estimated and verified for an actual biological network by means of an approximation with a Boolean function [11-13]. The origin of criticality is still unknown, but it is typically evaluated with respect to the fitness or function of an environment. In this sense, time constants of dynamics that select a network are much larger than time constants of dynamics within a network. The idea of SOC is characteristic of the comparative time constant that occurs between dynamics in and over systems. Thus, SOC is the intrasystemic mechanism used to both create and maintain a system. Therefore, SOC, instead of inter-systemic external selections, is a candidate for evolving a system [14-15].

The mechanism of SOC is not clear. Whereas actual biological systems are employed with local interactions, most SOC systems require global information for the system to operate [1, 2]. Then, the task at hand is to establish simple and general SOC frameworks in which an SOC mechanism can be implemented. Elementary cellular automata (ECA) [16-18] are hopeful candidates for this purpose.

Class 4 automata appear to be at the edge of chaos [7] and their computational universality is being investigated [18, 19]. As previous research has found phase transitions in ECA that have asynchronous updates [20-22], the mechanism of SOC can be implemented in the form of local interactions.

Here, we demonstrate that asynchronously tuned automata can implement SOC. We introduce asynchronous updating by updating orders defined by a bijective map, by defining local consistencies by adjunction in synchronous updates, and by generating asynchronous tuning and removing local inconsistencies of cellular automata.

Given a configuration consisting of $n$ cells, each of which is either a zero or one, an ECA can be described by the function $f:\{0, 1\}^3 \to \{0, 1\}$ which is called a local rule [16]. The time development of a configuration is defined by adapting a local rule to a configuration with periodic boundary conditions. Configuration in a time development is indexed by a natural number, $t$.

Asynchronous updating is introduced by an updating order defined by a bijective map, $Ord^t:\{1, 2, \ldots, n\} \to \{1, 2, \ldots, n\}$, which is randomly determined at each time step [23]; $Ord^t(k) \in \{1, 2, \ldots, n\} - \{Ord^t(1), Ord^t(2), \ldots, Ord^t(k-1)\}$ is chosen with equal probability. The $k$th cell in a configuration at the $t$th step is updated by the $Ord^t(k)$th order. If the state of the $k$th cell at the $t$th time step is represented by $a_k^t$, asynchronous updating is described by:

$$Ord^t(k-1) < Ord^t(k) < Ord^t(k+1) \Rightarrow a_k^{t+1} = f(a_{k-1}^{t+1}, a_k^t, a_{k+1}^t), \tag{1a}$$

$$Ord^t(k-1) > Ord^t(k) > Ord^t(k+1) \Rightarrow a_k^{t+1} = f(a_{k-1}^t, a_k^t, a_{k+1}^{t+1}), \tag{1b}$$

$$Ord^t(k-1) < Ord^t(k) > Ord^t(k+1) \Rightarrow a_k^{t+1} = f(a_{k-1}^{t+1}, a_k^t, a_{k+1}^{t+1}), \text{ and} \tag{1c}$$

$$Ord^t(k-1) > Ord^t(k) < Ord^t(k+1) \Rightarrow a_k^{t+1} = f(a_{k-1}^t, a_k^t, a_{k+1}^t). \tag{1d}$$

Note that (1d) means a local synchronous update.

Because asynchronous updating was randomly implemented, asynchronous ECA defined by equation (1) behaves like the Fates' asynchronous ECA, with updating having a probability $p$ and maintaining a state having a probability $1-p$ [20]. However, our asynchronous scheme can reveal more variety in a term consisting of actual local transitions. In the Fates' ECA, updated transitions that are inconsistent with a local rule are constrained under the condition $a_k^{t+1} = a_k^t$, whereas updated transitions that were inconsistent with a local rule in our asynchronous scheme were not constrained.

To define local consistencies in ECA, we introduced duality or adjunction into ECA. Any concept is described as a pair of intent and extent [24-26]. In set theory, a set $y$ is described as a collection of elements with respect to extent, and it is expressed as $x \in y$. A set $y$ is described as a characteristic that any elements in $y$ have with respect to intent, and it is expressed as $A(x)$. In this sense, intent is a replacement of extent and vice versa. When a pair of intent and extent values of a set of natural numbers smaller than $n$, ($A(x)$, $\{1, 2, …, n\}$), is compared with an even number, ($B(x)$, $\{2, 4, …, n\}$), the order of intents $A(x)<B(x)$ defined by the number of characters representing an intent is reversed by the order of the extent $\{1, 2, …, n\}>\{2, 4, …, n\}$, which are defined by cardinality. A series of intents is reversed by a series of extents in a term of the order. A pair of extents and intents, including their order (i.e., structure), is called *duality*.

Duality is described as an adjunction in category theory [27, 28]. A category *C* consists of objects *A*, *B*, … and arrows $f:A→B$, $g:B→C$, … that satisfy a particular condition. Between two categories, *C* and *D*, a functor, $F:C→D$, can be defined. Two functors, $F:C→D$ and $G:D→C$, are called adjoint functors if they satisfy the equivalence $F(C)→D⇔C→G(D)$, where *C* and *D* are objects in categories *C* and *D*, respectively. One-to-one correspondence between $F(C)→D$ and $C→G(D)$ is called adjunction.

If objects and arrows are defined by sets and maps, functors $A×(-)$ and $(-)^A$ are adjoint functors. If these adjoint functors are applied to $B→C$, a particular adjunction, $A×B→C⇔B→C^A$, is obtained, where $C^A$ represents a set of functions from *A* to *C*. Adjunction is found in the ECA local rule if $B = C = \mathbf{B} = \{0, 1\}$ and $A = \mathbf{B}×\mathbf{B}$. Adjunction results in $f_p$: $(\mathbf{B}×\mathbf{B})×\mathbf{B}→\mathbf{B}⇔f_a:\mathbf{B}→\mathbf{B}^{\mathbf{B}×\mathbf{B}}$. Given a local rule of ECA, $a_k^{t+1} = f(a_{k-1}^t, a_k^t, a_{k+1}^t)$ is expressed as a truth table:

$$000→d_0 \quad 001→d_1 \quad 010→d_2 \quad 011→d_3 \quad 100→d_4 \quad 101→d_5 \quad 110→d_6 \quad 111→d_7, \quad (2)$$

with $d_i \in \{0, 1\}$. A passive mode of the rule $f_p$: $(\mathbf{B}×\mathbf{B})×\mathbf{B}→\mathbf{B}$ is $f_p((a_{k-1}^t, a_{k+1}^t), a_k^t) = f(a_{k-1}^t, a_k^t, a_{k+1}^t)$. The truth table (2) is replaced by:

$$((0, 0), 0)→d_0, \quad ((0, 1), 0)→d_1, \quad ((1, 0), 0)→d_4, \quad ((1, 1), 0)→d_5 \text{ and} \quad (3a)$$
$$((0, 0), 1)→d_2, \quad ((0, 1), 1)→d_3, \quad ((1, 0), 1)→d_6, \quad ((1, 1), 1)→d_7. \quad (3b)$$

Table (2) is divided into (3a) and (3b) depending on the value of $a_k^t$. Here, it is interpreted that $a_k^t$ passively changes into $a_k^{t+1}$ using information regarding its nearest neighbors, $a_{k-1}^t$ and $a_{k+1}^t$.

An active mode of the rule is also defined if it is expressed as $f_a:\mathbf{B}→\mathbf{B}^{\mathbf{B}×\mathbf{B}}$ such that $f_a(a_k^t) = g \in \mathbf{B}^{\mathbf{B}×\mathbf{B}}$ with $g(a_{k-1}^t, a_{k+1}^t) = f(a_{k-1}^t, a_k^t, a_{k+1}^t)$ (i.e., $f_a(a_k^t)(a_{k-1}^t, a_{k+1}^t) = f(a_{k-1}^t, a_k^t, a_{k+1}^t)$). Truth Table (2) is replaced by:

$$0 → \{(0, 0)→d_0, \quad (0, 1)→d_1, \quad (1, 0)→d_4, \quad (1, 1)→d_5\} \text{ and} \quad (4a)$$
$$1 → \{(0, 0)→d_2, \quad (0, 1)→d_3, \quad (1, 0)→d_6, \quad (1, 1)→d_7\}. \quad (4b)$$

where a bracket represents a map in $\mathbf{B}^{\mathbf{B}×\mathbf{B}}$. Here, it is interpreted that $a_k^t$ actively changes into $a_k^{t+1}$ by itself through observing $a_{k-1}^t$ and $a_{k+1}^t$. A rearrangement of Table (2) results in Table (3) or (4). A passive mode of a local rule can be uniquely replaced by an active mode and vice versa. These modes are only different in their interpretation of a given local rule.

Now, we define asynchronously tuned automata in ECA. Although passive and active modes are not different with respect to their next state in synchronous updates, they can differ with respect to their next state in asynchronous updates. The active mode is applied to the 1d condition and the passive mode is applied to the 1a-c conditions; the passive mode is invariant through time and among all cells, and the active mode is locally tuned to be interpreted as the passive mode. "Asynchronously Tuned ECA" (AT_ECA) is defined by a given passive mode rule, active mode rule, and tuning rule. Given a local rule in equation 2 (i.e., a fixed set of $d_0$, $d_1$, …, $d_7$), a passive mode in AT_ECA is defined by $a_k^{t+1} = d_s$, where

$$Ord^t(k-1) < Ord^t(k) < Ord^t(k+1) \Rightarrow s = 4a_{k-1}^{t+1} + 2a_k^t + a_{k+1}^t; \quad (5a)$$
$$Ord^t(k-1) > Ord^t(k) > Ord^t(k+1) \Rightarrow s = 4a_{k-1}^t + 2a_k^t + a_{k+1}^{t+1}; \quad (5b)$$
$$Ord^t(k-1) < Ord^t(k) > Ord^t(k+1) \Rightarrow s = 4a_{k-1}^{t+1} + 2a_k^t + a_{k+1}^{t+1}. \quad (5c)$$

The active mode of the $k$th cell at the $t$th time step in the AT_ECA is defined by:

$$0 \rightarrow \{(0, 0) \rightarrow e_{0,k}^t \quad (0, 1) \rightarrow e_{1,k}^t, \quad (1, 0) \rightarrow e_{4,k}^t, \quad (1, 1) \rightarrow e_{5,k}^t\} \text{ and} \quad (6a)$$
$$1 \rightarrow \{(0, 0) \rightarrow e_{2,k}^t, \quad (0, 1) \rightarrow e_{3,k}^t, \quad (1, 0) \rightarrow e_{6,k}^t, \quad (1, 1) \rightarrow e_{7,k}^t\}. \quad (6b)$$

where $e_{s,k}^0 = d_s$ and $s = 0, 1, …, 7$. The active mode is applied only under the condition $Ord^t(k-1) > Ord^t(k) < Ord^t(k+1)$, which means that $a_k^{t+1} = f_a(a_k^t)(a_{k-1}^t, a_{k+1}^t) = e_{s,k}^t$ and $s = 4a_{k-1}^t + 2a_k^t + a_{k+1}^t$. Depending on the local order of updates, each cell is updated in either the active or passive mode. After updating, the tuning rule is applied to each cell. The tuning rule is defined by:

$$Ord^t(k-1) < Ord^t(k) < Ord^t(k+1) \Rightarrow e_{s,k}^{t+1} = d_0, \quad (7a)$$
$$Ord^t(k-1) > Ord^t(k) > Ord^t(k+1) \Rightarrow e_{s,k}^{t+1} = d_0, \text{ and} \quad (7b)$$
$$Ord^t(k-1) < Ord^t(k) > Ord^t(k+1) \Rightarrow e_{s,k}^{t+1} = a_k^{t+1}. \quad (7c)$$

If $Ord^t(k-1) > Ord^t(k) < Ord^t(k+1)$, no tuning happens at $k$th cell. This result means that if a cell is updated in the passive mode, then the active mode is reset in Equations 7a and 7b or it is tuned to be in the state produced by the passive mode in Equation 7c.

Given a rule number of ECA, a passive mode of the AT_ECA is uniquely determined and an active mode is temporally changed in rest and in tuning. Therefore, the AT_ECA is also coded by a rule number proposed by Wolfram [16].

One hundred and fifty-five of the 256 AT_ECA had class 4-like or cluster-like behaviors consisting of local periodic patterns and traveling waves among local patterns (Table 1). We called this class of ECA rules the "Critical Class". Fig. 1 illustrates some time developments of the AT_ECA with various Critical Class rules accompanied with time developments in synchronous updates. A rule that had class 3 when it was synchronously updated can have class 4-like patterns in asynchronously tuned updates. All class 3 rules in the synchronous updates belonged to the Critical Class. Class 3 rules and some rules showing class 1 or class 2 rules in synchronous updates can have a class 4-like pattern. It reveals that asynchronous tuning can drive a chaotic system (class 3) toward a more locally stable system and a stable system (class 1 or 2) toward a more chaotic system.

To estimate the behavioral changes that occurred from synchronous ECA to AT_ECA, metric entropy during time developments was measured [16, 29, 30]. When the frequency of a 4-bit configuration is interpreted as a probability of corresponding configurations, the entropy at each step can be obtained. Given a random configuration of $N$ cells, $M$ steps are discarded and metric entropies of $T$ steps are measured, and the mean and standard deviations of metric entropy over $T$ steps are obtained. Fig. 2a presents the standard deviations ($\sigma$) versus the mean entropies ($\mu$) for the local rules whose behaviors in the AT_ECA resembled the class 4-like pattern (i.e., the Critical Class); $N =$ 500, $M = 5$, and $T = 200$. Fig. 2b presents the same plot for time development produced by the same rules but for the synchronously updated system. Compared with patterns generated by the synchronously updated system, the patterns generated by the AT_ECA were characteristic of high $\sigma$ values independent of $\mu$ values.

One hundred of the 255 rules did not exhibit conspicuous differences between patterns generated by synchronous updates and patterns generated by the AT_ECA (Table 1, Fig. 3). We called this class of rules the "Ordinary Class". In Fig. 3, a curve obtained in Fig. 2a is superimposed onto a distribution of a pair of $\sigma$ and $\mu$ values for rules of the Ordinary Class. Note that pairs of $\sigma$ and $\mu$ for patterns generated by the synchronous updates and the AT_ECA are distributed under the curve. In other words, this curve revealed a class 4-like cluster pattern; a pattern whose $\sigma$ and $\mu$ pair does not exceed this curve is either the stable or chaotic pattern.

In a strict sense, the Critical Class can exhibit critical phenomena in the phase transition with respect to the power law in the decay of density. Focusing on cellular automata, phase transitions and/or critical phenomena are investigated in directed bond percolations [31-33]. Each one-dimensional site has two states: a media state, $r_k^t$ (open (1) or closed (0)), and a moisture state, $m_k^t$ (wet (1) or dry (0)). The moisture states of sites are updated according to the rule: $m_k^{t+1} = 1$ if $r_{k-1}^t = 1$ and $m_{k-1}^t = 1$ or $r_{k+1}^t = 1$ and $m_{k+1}^t = 1$; otherwise, $m_k^{t+1} = 0$, where $r_k^t$ can be open with probability $p$. This protocol mimics the fact that water drops percolate through porous (i.e., open) parts that are generated randomly in a media. This model illustrates a phase transition with respect to the probability of percolation, $Perc$, such that if water drops set in the $t = 0$ layer reach the $t = n$ layer,

*Perc* = 1; otherwise, *Perc* = 0. Actually, *Perc* = 0 if $p \leq p_c$ and *Perc* = 1 if $p_c < p$, where $p_c$ is a critical value. It is well known that the density of a water drop, $d(p_c, t)$, decays to zero and that this decrease follows a power law $d(p_c, t) \sim t^{-\delta}$; $\delta = 0.1595$.

The value of the exponent of decay, $\delta = 0.1595$, has a universality for other critical phenomena in cellular automata. As mentioned above, Fates' asynchronous ECA are updated with probability *p* or keep their previous state with probability 1-*p*. Some ECA updated according to this protocol exhibit class 4-like cluster patterns and phase transitions in a density term (i.e., a number of state 1 cells are normalized by the system size). When one ECA rule is chosen, a critical value for the probability of asynchronous updates is chosen, $p_c$, and a time development is generated. In this case, the decrease in density follows a power law with an exponent $\delta = 0.1595$ [22].

Even if a particular exponent of the power law has universality among a wide variety of phase transitions, the question regarding the origin of criticality remains unanswered as long as the power law is interpreted with respect to the phase transitions. In our system, most of the AT_ECA generating class 4-like patterns were autonomously tuned to be close to the critical point with $\delta = 0.1595$.

Fig. 4 presents density versus time in a log-log scale; the system size (*N*) was 1,000 and there were 100 trials (*K*). Each plot was obtained as the mean value of 100 trials, and each line corresponds to rules 22, 28, 54, 60, 70, 102, 124, 147, and 150. The black line illustrates that the power law decreases with $\delta = 0.1595$. This decrease in density coincides with the power law decrease of $\delta = 0.1595$ for rule 150 in the AT_ECA. Any other rules updated in the AT_ECA were also located close to the power law decay of $\delta = 0.1595$. Fig. 5 also presents the density versus time in a log-log scale, where rules 151, 156, 157, 182, 195, 198, 199, and 218 are updated in the AT_ECA. Other simulating conditions were the same as the conditions for the data presented in Fig. 4. The density decays of rules 151, 156, 157, and 182 were fitted to the line $\delta = 0.1595$. Other rules were also located close to the line.

These results indicate that the AT_ECA are autonomously tuned to a narrow band around the critical state. This behavior is a type of SOC. Whereas evolutionary biological systems are known to be stable at the edge of chaos, the origin of criticality is still unknown. Adaptive roles or fitness states that differ from the dynamics by which a system is generated and maintained in ontogeny might drive a system that is at the edge of chaos. In other words, because the time constant to maintain a system is different from the time constant to evolve a system, behavior following a particular order parameter is controlled by the former time constant and behavior tuning an order parameter is controlled by the latter. Systems have two types of time scales in this context, which suggests the presence of severe natural selection.

In contrast, SOC suggests that the critical state is easily achieved by the intrinsic mechanism of dynamics whose time constant is relevant for the dynamics to evolve a system. In a Bak and

Sneppen model for an evolutionary ecosystem, dynamics to maintain a system coincide with dynamics to evolve a system. Thus, a critical state is achieved as a steady state. Dynamics to evolve a system can consider an entire system as an element. On the other hand, dynamics to maintain a system can consider the component of a system as an element. In fact, two types of dynamics are mixed in the Bak and Sneppen model, and the property as a whole is linked with a local property. The species with the lowest fitness (global property) is abandoned, along with the nearest neighboring species (local property) in terms of the food chain. An adequate balance of the local property with the global property autonomously drives the system to a critical state.

The AT_ECA is a type of SOC but it might yield another mechanism for SOC. The Bak and Sneppen model and other SOC models can balance the local property with the global property. Because these models have common mechanisms by which the driving force toward global consistency is perturbed by local interactions, consistency as an entire system is assumed. In contrast, the AT_ECA never implements global consistency but instead implements a dynamics between consistency and inconsistency, where consistency is defined as a consistent transition between the active and passive modes. The AT_ECA has no global information, such as the species with the lowest fitness in an ecosystem. Active and passive modes are equivalent to each other in the form of adjunction as long as the system is updated synchronously. Because the adjunction in synchronous updates can yield consistency in a system, asynchronous updates can derive inconsistency and tuning and resetting can entail methods to remove inconsistencies. Regardless of the manner in which tuning and resetting is actualized, consistency cannot be achieved because of the asynchronous updates; thus, another inconsistency is generated. Therefore, the endeavor to remove inconsistencies is successively continued, which is the main mechanism of SOC.

Mathematical duality, called adjunction, is typically used to construct a simple conjugate pair. Even if an adjunction pair of equations is equivalent to each other, one is complex to solve and the other can be solved easily. In this case, a simple conjugate is constructed to solve the equation. However, duality itself is not typically used. In our findings, adjunction played an essential role in generating and removing inconsistencies derived by asynchronous updates, which can lead to SOC. Adjunction coupled with asynchronous updates may highlight a general mechanism of SOC and/or critical phenomena in biological systems.

[1]    P. Bak, C. Tang, and K. Wiesenfeld, Phys. Rev. Lett. **59**, 381 (1987).
[2]    P. Bak and K. Sneppen, Phys. Rev. Lett. **71**, 4083 (1993).
[3]    P. Bak and C. Tang, J. Geol. Res. **94**, 15635 (1989).
[4]    D. R. Chialvo and P. Bak, Neuroscience **90**, 1137 (1999).
[5]    P. Bak, Phys. Rev. E **63**, 031912 (2001).
[6]    S. Bornholdf and T. Röhl, Phys. Rev. E **67**, 066118 (2003).

[7] C. G. Langton, Physica D **42**, 12 (1990).

[8] S. A. Kauffman and S. Johnsen, J. Theor. Biol. **149**, 467 (1991).

[9] J. Garcia, M. Gomos, T. Jyh, T. Ren, and T. Sales, Phys. Rev. E **48**, 3345 (1993).

[10] N. Bertschinger and T. Natschläger, Neur. Comp. **16**, 1413 (2006).

[11] S. A. Kauffman, J. Theor. Biol. **22**, 437 (1969).

[12] T. Rohlf, N. Gulbahce, and C. Teuscher, Phys. Rev. Lett. **99**, 248701 (2007).

[13] T. Rohlf, EPL **84**, 10004 (2008).

[14] K. Ito and Y.–P. Gunji, Biosystems **26**, 135 (1992).

[15] Y.-P. Gunji and M. Kamiura, Physica D **198**, 72 (2004).

[16] S. Wolfram, Rev. Mod. Phys. **55**, 601 (1983).

[17] S. Wolfram, Physica D **10**, 1 (1984).

[18] S. Wolfram, *A New Kind of Science* (Wolframscience.com, 2002).

[19] M. Cook, Complex Systems **15**, 1 (2004).

[20] T. Ingerson and R. Buvel, Physica D **10**, 59 (1984).

[21] N. Fatès, É. Thierry, M. Morvan, and N. Schabanel, Theor. Comp. Phys. **362**, 1 (2006).

[22] N. Fatès, *Asynchronism Induces Second Order Phase Transitions in Elementary Cellular Automata* (http://arxiv.org/pdf/nlin/0703044v2.pdf, 2008).

[23] Y. Gunji, BioSystems **23**, 317 (1990).

[24] J. Barwise, D. Gabbay, and C. Hartonas, Logic J IGPL **3**, 7 (1995).

[25] J. Barwise and J. Seligman, *Information flow: the logic of distributed systems* (Cambridge University Press, Cambridge, UK, 1997).

[26] B. Ganter, R. Wille, and C. Franzke, *Formal Concept Analysis: Mathematical Foundations* (Springer-Verlag, New York, USA, 1997).

[27] S. MacLane, *Categories for the Working Mathematicians* (Springer-Verlag, NewYork, USA, 1998).

[28] S. Awodey, *Cateogory Theory* (Oxford University Press, Oxford, UK, 2010).

[29] A. Wuenshe, Complexity **4**, 47 (1999).

[30] D. Uragami and Y.-P. Gunji, Physica D **237**, 187 (2008).

[31] E. Domany and W. Kinzel, Phys. Rev. Lett. **53**, 311 (1984).

[32] N. Fatès, Complex Systems **16**, 1 (2005).

[33] H. Hinrichsen, Advances in Physics **49**, 815 (2000).

Table 1. Classification of Critical Class (Bold) and Ordinary Class (Non-Bold) rules of the AT_ECA. Numbers represent Wolframs's rule number of ECA. A set of rules included in a bracket is an equivalence class under the symmetries 0/1 and left/right. Note that rules in an equivalent class can belong to different classes. Specifically, rule 160 exhibits class 1 behavior for the Ordinary Class and rule 260 exhibits class 4 behavior for the Critical Class.

| | | | |
|---|---|---|---|
| 0 (255), | 1 (127), | **2 (191 16 247)**, | 3 (63 17 119), |
| 4 (223), | 5 (95), | **6 (159 20 215)**, | **7 (31 21 87)** |
| 8 (239 64 253), | 9 (111 65 125), | **10 (175 80 245)**, | 11 (47 81 117), |
| 12 (207 68 221), | 13 (79 69 93), | **14 (143 84 213)**, | **15 (85)**, |
| **18 (183)**, | **19 (55)**, | **22 (151)**, | **23**, |
| **24 (231 66 189)**, | 25 (103 67 61), | **26 (167 82 181)**, | 27 (39 83 53), |
| **28 (199 70 157)**, | 29 (71), | **30 (135 86 149)**, | 32 (251), |
| 33 (123), | **34 (187 48 243)**, | 35 (59 49 115), | 36 (219), |
| 37 (91), | **38 (155 52 211)**, | 40 (235 96 249), | 41 (107 97 121), |
| **42 (171 112 241)**, | 43 (113), | 44 (203 100 217), | **45 (75 101 89)**, |
| **46 (139 116 209)**, | **50 (179)**, | 51, | **54 (147)**, |
| **56 (227 98 185)**, | 57 (99), | **58 (163 114 177)**, | **60 (195 102 153)**, |
| **62 (131 118 145)**, | 72 (237), | **73** (109), | **74 (173 88 229)**, |
| 76 (205 76 205), | 77, | **78 (141 92 197)**, | **90 (165)**, |
| **94 (133)**, | 104 (233), | 105, | **106 (169 120 225)**, |
| 108 (201), | **110 (137 124 193)**, | **122 (161)**, | **126 (129)**, |
| 128 (254), | 130 (190 144 246), | 132 (222), | 134 (158 148 214), |
| 136 (**238** 192 **252**), | **138 (174 208 244)**, | 140 (**206** 196 **220**), | **142 (212)**, |
| **146 (182)**, | **150**, | **152 (230 194 188)**, | **154 (166 210 180)**, |
| **156 (198)**, | 160 (**250**), | **162 (186 176 242)**, | 164 (**218**), |
| 168 (**234** 224 **248**), | **170 (240)**, | 172 (**202** 228 **216**), | 178, |
| **184 (226)**, | 200 (236), | 204, | 232 |

Figure Legends

Figure 1. A pair of patterns generated by ECA (right) and the AT_ECA (left). Each pattern proceeds vertically. The accompanied number represents a Wolfram's rule number for ECA.

Figure 2. The standard deviation of the spatial metric entropy versus the mean entropy for the pattern generated by ECA (right) and the AT_ECA (left) for the Critical Class rules. The curve is fitted to the AT_ECA plots. Each cross represents each rule. Pairs of $\sigma$ and $\mu$ are distributed along a curve that is convex toward the top of the figure and is drawn by a broken line.

Figure 3. The standard deviation of the spatial metric entropy versus the mean entropy for the pattern generated by ECA (right) and the AT_ECA (left) for the Ordinary Class rules. The curve fitted for the AT_ECA plot of the Critical Class is superimposed.

Figure 4. Density versus time on a log-log scale for some Critical Class rules (rule numbers are represented in the graph). Plots for rule 150 are fit to a line with an exponent of -0.1595.

Figure 5. Density versus time in a log-log scale for some Critical Class rules (rule numbers are represented in the graph). Plots for rules 151, 156, 157, and 182 are fit to a line with an exponent of -0.1595.

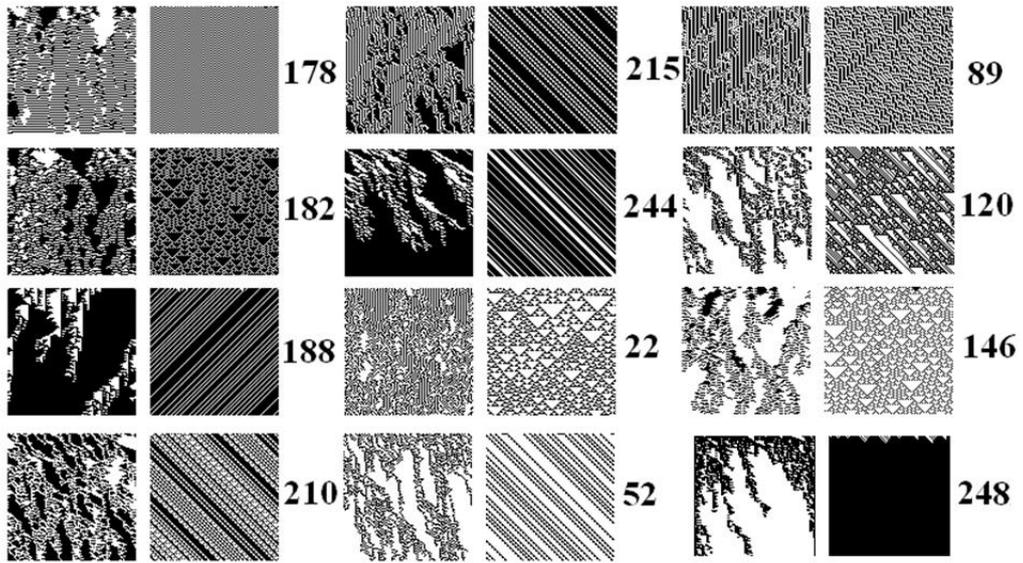

Fig. 1

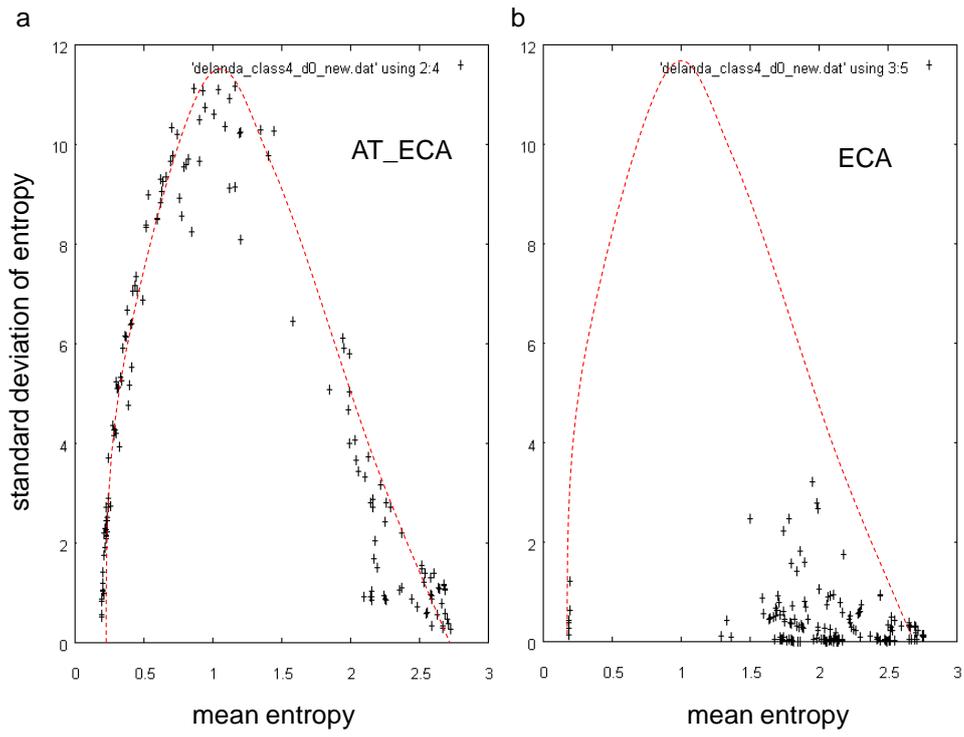

Fig. 2

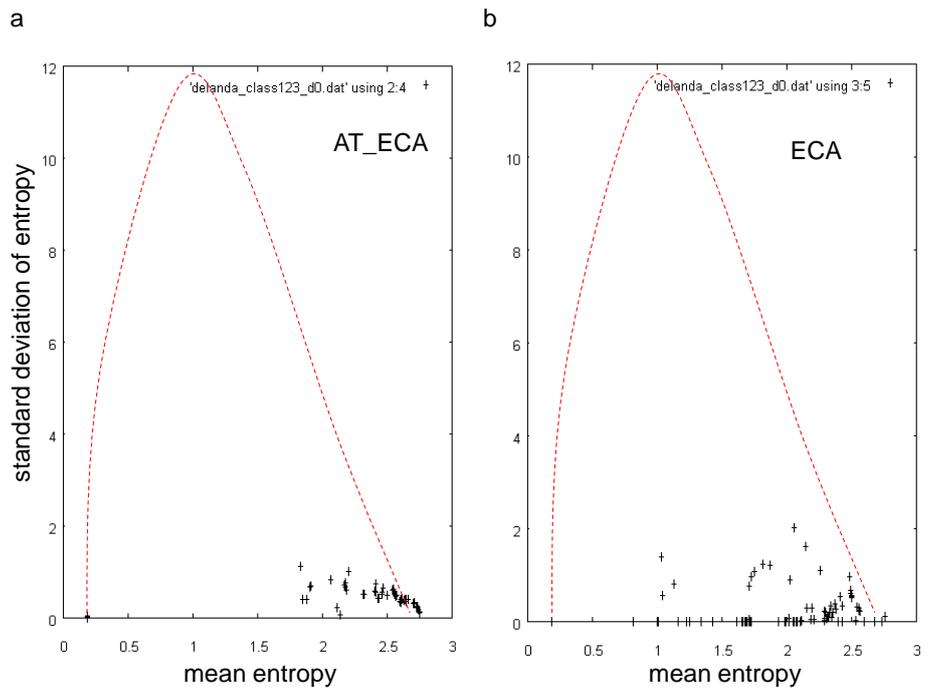

Fig. 3

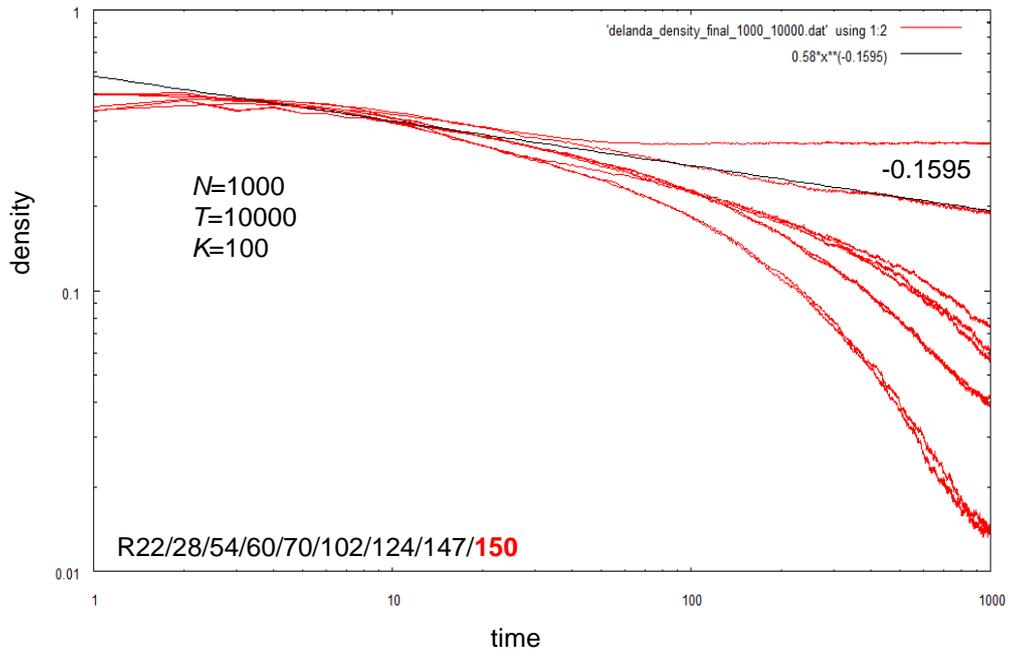

Fig. 4

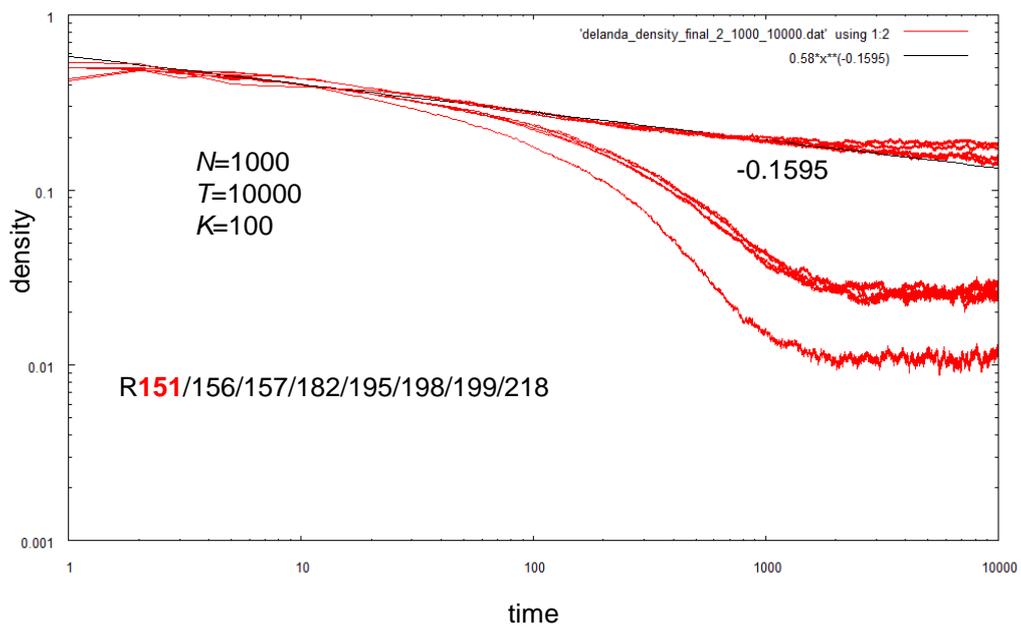

Fig. 5